\documentclass[%
twocolumn, prl, aps, superscriptaddress, longbibliography,showpacs,amsmath,amssymb,floatfix
]{revtex4-2}

\usepackage{changes}
\usepackage{graphicx}
\usepackage{dcolumn}
\usepackage{bm}
\usepackage{graphicx}                                   
\usepackage{amssymb}
\newtheorem{theorem}{Theorem}
\usepackage{amsmath}
\newtheorem{corollary}{Corollary}
\usepackage{epsfig}
\usepackage{xcolor}
\usepackage{tabu}
\usepackage{mathtools}
\usepackage[colorlinks,linkcolor=blue,anchorcolor=blue,citecolor=blue,urlcolor=blue]{hyperref}
\usepackage{physics}
\usepackage{float}
\usepackage{diagbox}
\usepackage{inputenc}

\begin{document}
\title{Generating function for Hermitian and non-Hermitian models}
\author{Hua-Yu Bai}
\affiliation{Laboratory of Quantum Information, University of Science and Technology of China, Hefei 230026, China}
\affiliation{Anhui Province Key Laboratory of Quantum Network, University of Science and Technology of China, Hefei, Anhui, 230026, China}
\affiliation{CAS Center For Excellence in Quantum Information and Quantum Physics, University of Science and Technology of China, Hefei 230026, China}

\author{Yang Chen}
\email{chenyang@ustc.edu.cn}
\affiliation{Laboratory of Quantum Information, University of Science and Technology of China, Hefei 230026, China}
\affiliation{Anhui Province Key Laboratory of Quantum Network, University of Science and Technology of China, Hefei, Anhui, 230026, China}
\affiliation{CAS Center For Excellence in Quantum Information and Quantum Physics, University of Science and Technology of China, Hefei 230026, China}

\author{Guang-Can Guo}
\affiliation{Laboratory of Quantum Information, University of Science and Technology of China, Hefei 230026, China}
\affiliation{Anhui Province Key Laboratory of Quantum Network, University of Science and Technology of China, Hefei, Anhui, 230026, China}
\affiliation{CAS Center For Excellence in Quantum Information and Quantum Physics, University of Science and Technology of China, Hefei 230026, China}

\author{Ming Gong}
\email{gongm@ustc.edu.cn}
\affiliation{Laboratory of Quantum Information, University of Science and Technology of China, Hefei 230026, China}
\affiliation{Anhui Province Key Laboratory of Quantum Network, University of Science and Technology of China, Hefei, Anhui, 230026, China}
\affiliation{CAS Center For Excellence in Quantum Information and Quantum Physics, University of Science and Technology of China, Hefei 230026, China}

\author{Xi-Feng Ren}%
\email{renxf@ustc.edu.cn}
\affiliation{Laboratory of Quantum Information, University of Science and Technology of China, Hefei 230026, China}
\affiliation{Anhui Province Key Laboratory of Quantum Network, University of Science and Technology of China, Hefei, Anhui, 230026, China}
\affiliation{CAS Center For Excellence in Quantum Information and Quantum Physics, University of Science and Technology of China, Hefei 230026, China}

\begin{abstract}
It is well known that Hermitian and non-Hermitian models exhibit distinct physics and require different theoretical tools. In this work, we propose a unified generating-function framework for both classes with generic boundary conditions and local impurities. Within this framework, any finite lattice model can be mapped to a generating function of the form $\mathcal{G}(z)=P(z)/Q(z)$, where $Q(z)$ and $P(z)$ denote the bulk recurrence relation and boundary terms or impurities, respectively. The problem of solving for eigenstates reduces to a simple criterion based on the cancellation of zeros of $Q(z)$ and $P(z)$. Applying this method to the Hatano–Nelson (HN) model, we show how boundary conditions and impurities determine the location of the zeros, thereby demonstrating the boundary sensitivity of non-Hermitian systems. We further investigate topological edge states in the non-Hermitian Su-Schrieffer-Heeger (SSH) model and identify its topological phase transition. Inspired by generating-function techniques widely used in discrete mathematics, particularly in the study of the Fibonacci sequence, our results establish a direct connection between non-Hermitian physics and recurrence relations, providing a new perspective for analyzing non-Hermitian systems and exploring their connections with discrete mathematical structures. 
\end{abstract}

\maketitle

The generating function approach has long played an important role in number theory,  combinatorics, and probability theory \cite{herbert1994generating,andrews1984,flajolet1990singularity}. It has been widely applied to the study of Diophantine equations and integer partitions \cite{andrews2004integer,stanley2015catalan,comtet2012advanced,mireille2006polynomial}, as well as to the analysis of random walks via the kernel method \cite{cyril2002basic,helmut2004kernel,flajolet2009analytic,mireille2005walks,bona2006walk}. In the kernel method, the central idea is to determine the generating function by requiring a zero of the numerator to coincide with a zero of the denominator. Closely related ideas appear in the functional Bethe Ansatz and Baxter's TQ method \cite{baxter2016exactly,sklyanin1988boundary,rafael2003bethe,nepomechie2013inhomogeneous}, where the transfer-matrix eigenvalue is required to be analytic in the complex spectral parameter.

\begin{figure}[h]
\includegraphics[width=0.49\textwidth]{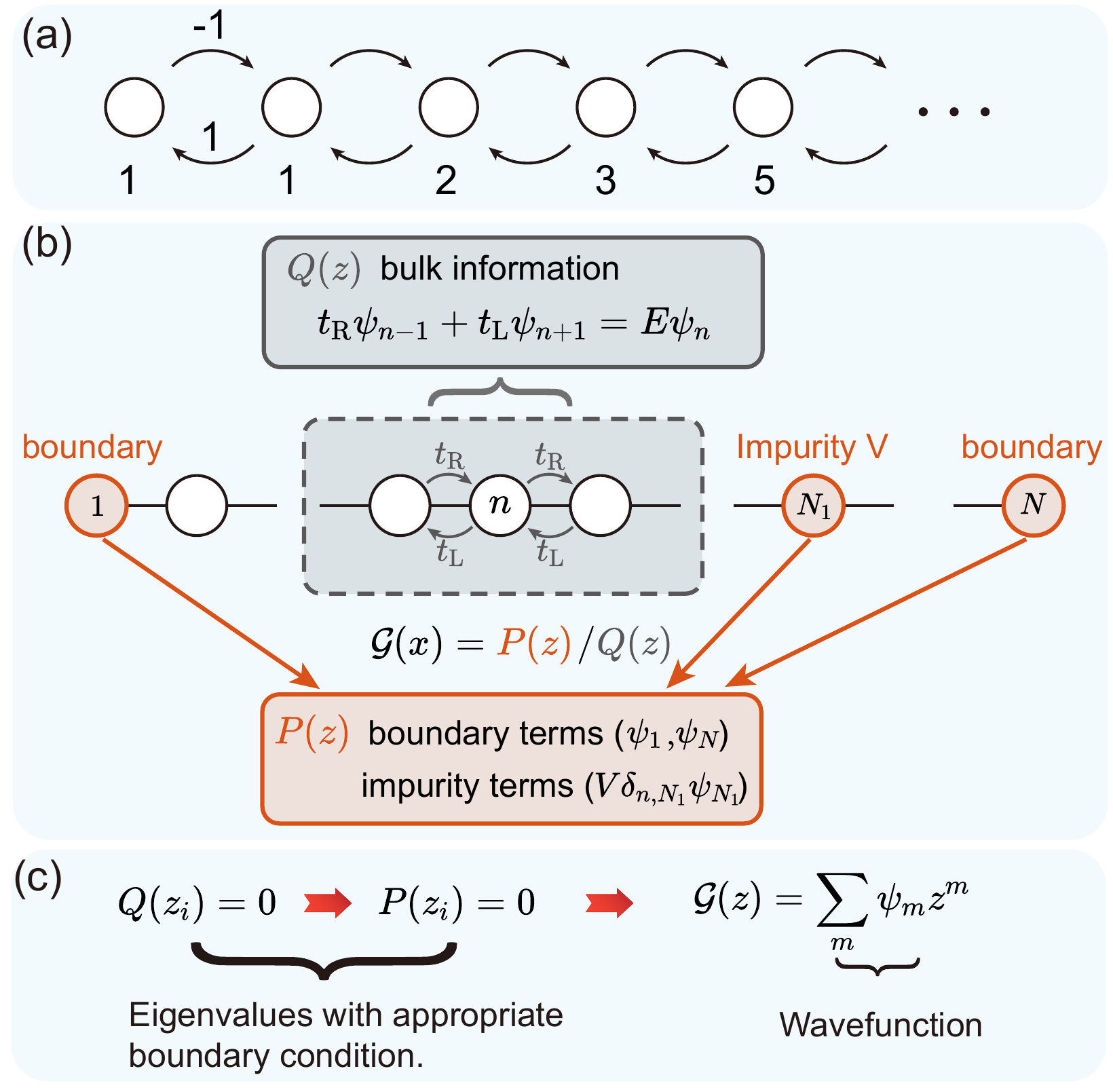}
\caption{(a) Diagram illustrating the recurrence relation of the Fibonacci sequence in terms of  a non-Hermitian model.  (b) Representation of the unified approach based on generating function for one-dimensional hopping models $\mathcal{G}(z)=P(z)/Q(z)$. (c) The criterion of cancellation of zeros,  where admissible eigenvalues are selected by pairing the zeros, and then we get the wavefunction.}
\label{fig1}
\end{figure}

In this manuscript, from the perspective of generating function, we propose a unified framework for solving wavefunctions in both Hermitian and non-Hermitian models, incorporating various boundary conditions and impurities on an equal footing. In non-Hermitian models \cite{yuto2020non,hodaei2017enhanced,brody2013biorthogonal,esaki2011edge,shen2018topological,solinas2026biorthogonal}, major research focuses have included topology \cite{bergholtz2021exceptional,kawabata2019symmetry,borgnia2020non,helbig2020generalized,xiao2020non,wang2021generating},  non-Bloch bands 
\cite{yokomizo2019non,hu2024non,longhi2020non},  exceptional points \cite{bender1998real,bender2007making,rüter2010observation,feng2013experimental,peng2014loss,el2018non,liu2021exactm,parto2025enhanced}, and disorder \cite{li2025universal,rivero2023robust,longhi2025erratic,sun2021geometric}. In particular, under the non-Hermitian skin effect \cite{song2019non,yi2020non,zhang2022universal,li2022gain,lee2019hybrid,zhu2022hybrid,longhi2022self,weidemann2020topological,liang2022dynamic,zhang2021acoustic,zhang2021observation,gao2022anomalous,gu2022transient,sun2024photonic,lin2024observation,xue2024self,zhao2025two,zhou2023observation,zou2021observation}, boundary conditions have a significant impact on wavefunctions. Within our framework, we derive the eigenvalues and wavefunctions for systems under open boundary conditions (OBC) and periodic boundary conditions (PBC),  including cases with impurities.  This approach provides a clear perspective on several central results in one-dimensional systems:
(I) It reproduces the generalized Brillouin zone (GBZ) \cite{yokomizo2019non,yao2018edge,Xiong2018bulk,wu2022connections}  in non-Hermitian systems and clarifies their boundary sensitivity;
(II)  It determines impurity-bound states and the associated decay rates of wavefunctions \cite{li2021impurity,guo2023accumulation,liu2021exact,roccati2021non,lang2018effects}; (III) It diagnoses the existence of topological edge states in the non-Hermitian SSH model \cite{yao2018edge,lieu2018topological}.
 
Our conceptual breakthrough lies in a reexamination of the iterative method across various fields while establishing some subtle connections among them. For the Fibonacci number $b_n=b_{n-1}+b_{n-2}$ for $n\ge 2$ with $b_0=0$, $b_1=1$, we define the generating function $\mathcal{G}(z)=\sum_{m\ge 0}b_m z^m$.  
It follows that $ \mathcal{G}(z)=z+z \mathcal{G}(z)+z^2 \mathcal{G}(z)$,  and hence $\mathcal{G}(z) = z/(1-z-z^2) = P(z)/ Q(z)$, where $P(z) =z$ and $Q(z) = 1 -z -z^2$. We obtain 
\begin{equation}
    \mathcal{G}(z)=\frac{1}{z_1-z_2}\sum_{m\ge0}(z_1^{-m}z^m-z_2^{-m}z^m), 
    \label{eq-wf-fibonacci}
\end{equation}
yielding $b_m=\frac{1}{z_1-z_2}(z_1^{-m}-z_2^{-m})$, where $z_1$ and $z_2$ are zeros of $Q(z)$ with $z_{1,2}=\frac{-1\pm \sqrt{5}}{2}$. When an impurity term $\delta_{n,10}$ is added, so that $b_n=b_{n-1}+b_{n-2}+\delta_{n,10}$,  we find $ \mathcal{G}(z)=z(1 + z^9)/(1 - z - z^2)$. Here, $P(z)$ becomes $z(1 +z^9)$ while $Q(z)$ remains unchanged. 


This idea is stimulating because the Fibonacci sequence can be mapped to the non-Hermitian Hamiltonian $ H=\sum_n  a_n^\dagger a_{n+1}- a_{n+1}^\dagger a_n$, with a semi-infinite boundary condition ($\psi_0=0$) and $E=1$ [see Fig. \ref{fig1}(a)].  The solution takes the form $\psi = \sum_{n\ge 1} \psi_n |n\rangle$, satisfying  $H\psi = E \psi$. When $E \ne 1$, the recurrence becomes $\psi_{n+1} = E \psi_n + \psi_{n-1}$, yielding $\psi_m=\frac{\psi_1}{z_1-z_2}(z_1^{-m}-z_2^{-m})$, where $z_{1,2}$ are the roots of $z^2+E z-1=0$. For the more general interaction equation $b_n = c_1 b_{n-1} + c_2 b_{n-2} + \cdots + c_k b_{n-k}$ for $n\ge k$, $\mathcal{G}(z) = P(z)/(1 - c_1 z - c_2 z^2 - \cdots -c_k z^k)$, where $P(z)$ is determined by the initial conditions $b_0$, $b_1$, $\cdots$, $b_{k-1}$. These initial terms are related to the corollary presented later.  This perspective can greatly broaden our understanding of non-Hermitian physics by establishing a linkage between the two distinct fields. 



In our formulation, the boundary conditions and the impurities are encoded in the numerator $P(z)$, while $Q(z)$  arises from the recurrence relation [Fig. \ref{fig1}(b)]. Our main result can be summarized in terms of the following theorem.
\begin{theorem}\label{thm:pole-cancel}
Consider a generating function of a physical system expressed in the form $\mathcal{G}(z) = P(z)/Q(z)$,  where \(P(z)\) and \(Q(z)\) are  polynomials in \(z\). If the wavefunction expansion of the system requires that \(\mathcal{G}(z)\) be expressible, under certain boundary conditions (open or periodic), as a finite-degree polynomial, then a necessary (and often sufficient) condition is: (A) All zeros of \(Q(z)\) must coincide with zeros of \(P(z)\), so that the poles of \(\mathcal{G}(z)\) at these points are canceled; (B) This cancellation condition yields equations for the physical parameters (propagating factors, energy, etc.), whose solutions correspond to the allowed eigenstates of the system. 
\end{theorem}
Based on this theorem and the definition of $\mathcal{G}(z)$, we have the following 
corollaries \cite{chenintroduction,suppbai}.
\begin{corollary}\label{def-corollary}
Given $\mathcal{G}(z)=\sum_{m\ge 0}b_m z^m$ and $h>0$, we have
\begin{equation}
    \sum_{m\ge 0}b_{m+h} z^m=\frac{\mathcal{G}(z)-b_{0}-\dots-b_{h-1} z^{h-1}}{z^{h}}.
\end{equation}
\end{corollary}
\begin{corollary}\label{def-corollary2}
Given that $p_m$ is a polynomial in $m$, we have
\begin{equation}
    p(z D) \mathcal{G}(z)= \sum_{m\ge 0} p_m b_{m} z^m,
\end{equation}
where $D = {d \over dz}$. In particular, when $p_m = e^{-\lambda m}$, we have 
\begin{equation}
    \mathcal{G}(e^{-\lambda} z)=   \sum_{m\ge 0} e^{-\lambda m} b_m  z^m.
\end{equation}
\end{corollary}

\begin{figure}[b]
\includegraphics[width=0.49\textwidth]{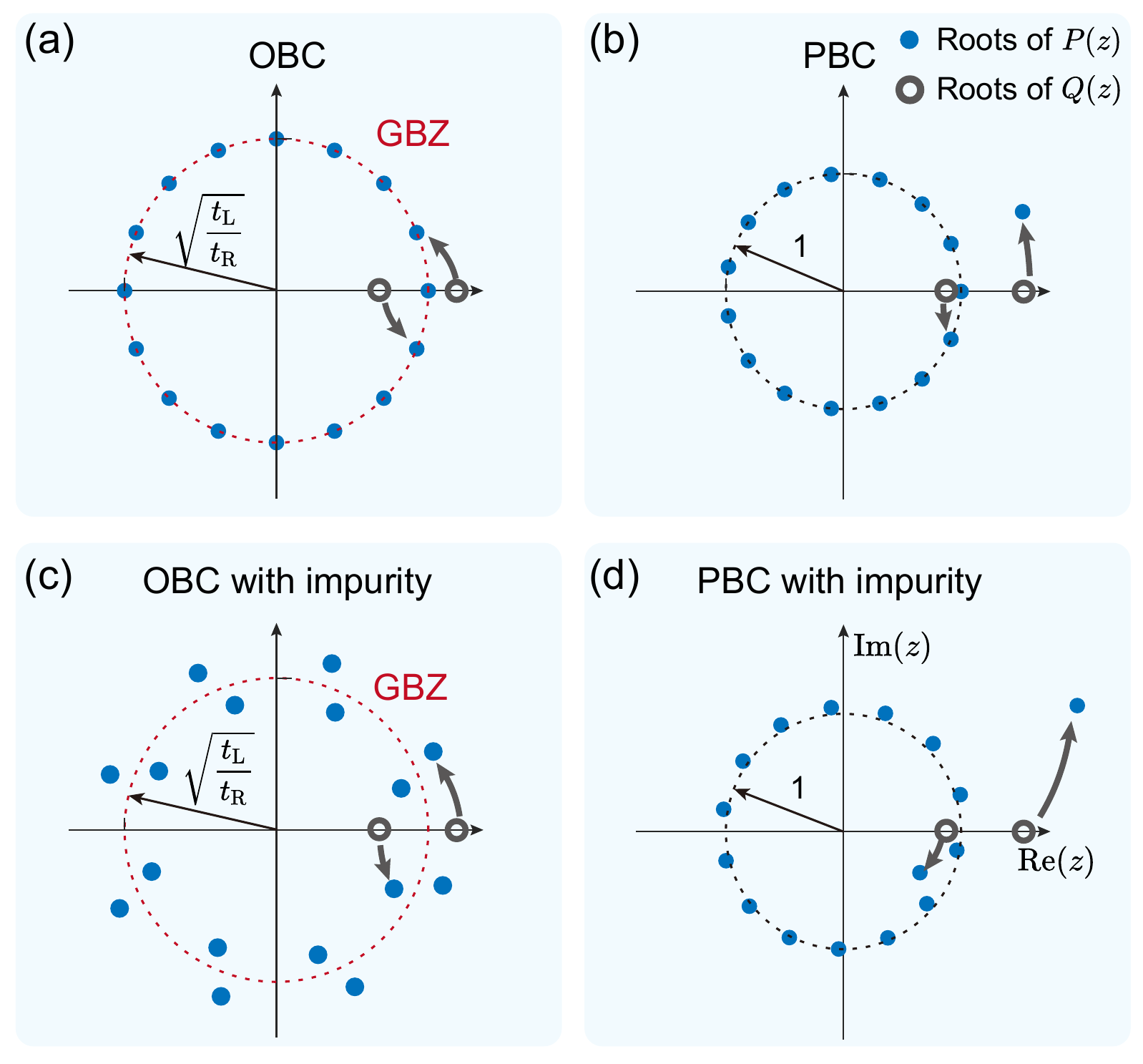}
\caption{Distribution of the zeros of $P(z)$ excluding $z=0$ in the complex plane with different boundary settings for the HN model. (a) OBC, (b) PBC, (c) OBC with an impurity, (d) PBC with an impurity. Blue dots mark zeros of $P(z)$, grey circles mark zeros of $Q(z)$. Arrows indicate the cancellation of zeros of $Q(z)$ with zeros of $P(z)$, which  correspond to the admissible states. The red dashed circle represents the GBZ condition $|z|=\sqrt{t_\text{L}/t_\text{R}}$.
}
\label{fig2}
\end{figure}

The above corollaries also establish a transformation of $P(z)$ and $Q(z)$ accordingly. Corollary \ref{def-corollary}  will be applied to the derivation of $\mathcal{G}(z)$ and will provide boundary terms, showing the change of $\mathcal{G}(z)$ by removing some boundary sites. Corollary \ref{def-corollary2} establishes a relation between two Hamiltonians under the transformations $a_m \to p_m^{-1} a_m$ and $a_m^\dagger \to p_m  a_m^\dagger$,  under which $t_m^\text{L} \rightarrow p_m p_{m+1}^{-1} t_m^\text{L}$ and $t_m^\text{R} \rightarrow p_{m+1} p_m^{-1} t_m^\text{R}$. The choice $p_m = 
e^{-\lambda m}$ corresponds precisely to the similarity transformation between Hermitian and non-Hermitian models. $P(z)$ and $Q(z)$ are changed to $P(e^{-\lambda}z)$ and $Q(e^{-\lambda}z)$, which means that their common zeros are rescaled by $e^\lambda$.  This method can be used to  eliminate nonuniform coupling phases and solve the non-Hermitian model with disorder \cite{longhi2025erratic}.

We now discuss the symmetries of $\mathcal{G}(z)$. When $b_n=\pm b_{N+1-n}$ for $n= 1,2,\dots,N$ and $\mathcal{G}(z)=\sum_{m=1}^{N}b_m z^m$, we have $ \mathcal{G}(z)=\pm z^{N+1} \mathcal{G}(1/z)$. This relation will be illustrated in the Hermitian case of the HN model. The manifestation of chiral symmetry in the generating function will be discussed later in the SSH model.

Based on the position of the poles, we can determine the eigenvalues and the wavefunctions simultaneously for Hermitian and non-Hermitian models, which can be treated on an equal footing. In this paper, we demonstrate these results using several concrete models. In addition, we can decompose $\mathcal{G}(z)$  into left boundary, right boundary, and impurity contributions and expand each term as an infinite series. One finds that the wavefunction amplitudes outside the chain  ($m \le 0$ and $m \ge N+1$) are proportional to $P(z_1)$ and $P(z_2)$ respectively. Consequently, $P(z_{1,2})=0$ ensures that all amplitudes outside the chain originating from these contributions cancel out \cite{suppbai}.

{\it (I) $\mathcal{G}(z)$ in the HN model with OBC and PBC}:  We aim to show that the generating function $\mathcal{G}(z)$ demonstrates the sensitivity of non-Hermitian systems to various boundary conditions. For the  HN model \cite{hatano1996localization,hatano1997vortex,hatano1998non,maddi2024exact,ochkan2024non,Orsel2025Giant,gliozzi2024many} of  $N$ sites,  the Hamiltonian is 
\begin{equation}
    H=\sum_n t_\text{L} a_n^\dagger a_{n+1}+ t_\text{R} a_{n+1}^\dagger a_n.\label{HNHamiltonian}
\end{equation}
The recurrence relation is $t_\text{R}\psi_{n-1}+t_\text{L}\psi_{n+1}=E\psi_n$, and we define $\mathcal{G}(z)=\sum_{m=1}^{N}\psi_m z^m$. 

Our major results are summarized as follows: (1) For the case with OBC, we impose $\psi_0=0$ and $\psi_{N+1}=0$. Using the finite-length analogue of Corollary \ref{def-corollary}, we obtain
\begin{equation}
    \mathcal{G}(z)=\frac{z(t_\text{L} \psi_1+t_\text{R}\psi_N z^{N+1})}{t_\text{R}z^2-E z+t_\text{L}}. \label{GOBC}
\end{equation}

Therefore, $P(z)=t_\text{L}\psi_1 z+t_\text{R} \psi_N  z^{N+2}$ and $Q(z)=t_\text{R} z^2-E z+t_\text{L}$. The zeros of $Q(z)$ are denoted as $z_{1,2}$. All the nonzero zeros of $P(z)$ lie on the circle of radius $|\frac{t_\text{L}\psi_1}{t_\text{R}\psi_N}|^\frac{1}{N+1}$ [Fig. \ref{fig2}(a)]. Theorem 1(A) implies that $z_1$ and $z_2$ are each paired with one of these zeros, yielding $P(z_i)=0 $ ($i=1,2$), from which we obtain $z_1^{N+1}=z_2^{N+1}$, namely the GBZ condition \cite{yokomizo2019non,yang2020non,gong2018topological,zhang2020correspondence}. Then using $z_1 z_2=t_\text{L}/t_\text{R}$,   we obtain  $z_1=\sqrt{t_\text{L}/t_\text{R}}e^{i\theta \ell}$ and $z_2 = \sqrt{t_\text{L}/t_\text{R}}e^{-i\theta \ell}$, where $\theta=\frac{\pi}{N+1}$ and  $ \ell\in \{1,\dots,N\}$.  We can obtain the eigenvalues as  $E=t_\text{R}(z_1+z_2)=2\sqrt{t_\text{R}t_\text{L}}\cos(\ell \theta)$, which yields Theorem 1(B). This process is schematically shown in Fig. \ref{fig1}(c). Moreover, we have  $\psi_N=\pm (t_\text{R}/t_\text{L})^{\frac{N-1}{2}}\psi_1$, yielding $\mathcal{G}(z)=\pm z^{N+1} \mathcal{G}(1/z)$ under the condition $t_\text{L}=t_\text{R}$, which corresponds to the reflection symmetry of the wavefunctions $\psi_m=\pm \psi_{N+1-m}$ \cite{suppbai}.

Reading off the coefficient of $z^m$ gives the wavefunction
\begin{equation}
    \psi_m=-\frac{t_\text{L} \psi_1 }{t_\text{R}(z_1-z_2)}(z_1^{-m}-z_2^{-m}),
\end{equation}
using $\mathcal{G}(z) = \sum_{m=1}^{N} \psi_m z^m$; see Eq. \ref{eq-wf-fibonacci} and Fig. \ref{fig1}(c). Thus the information of the zeros determines the eigenvalues, while the polynomial coefficients determine the corresponding wavefunctions.

We can characterize the skin modes using the winding number $W(E)=-\frac{1}{2\pi i}\oint_{|z|=1} \frac{\mathrm{d}}{\mathrm{d}z} \log(Q(z)/z)\mathrm{d}z$. $W(E)=+1(-1)$ indicates a leftward (rightward) skin effect \cite{suppbai}. 

(2) For the case with PBC, with $\psi_n=\psi_{n+N}$, we have
\begin{equation}
    \mathcal{G}(z)=\frac{z(t_\text{L}\psi_1-z t_\text{R}\psi_N)(1-z^N)}{t_\text{R}z^2 -E z+t_\text{L}}. \label{GPBC}
\end{equation}
Theorem 1(A) yields $t_\text{L}\psi_1-z_1 t_\text{R}\psi_N=0$ and  $1-z_2^N=0$. As shown in Fig. \ref{fig2}(b),  the zeros of $P(z)$ lie on the unit circle, together with an additional zero at $z=t_\text{L}\psi_1/(t_\text{R}\psi_N)$. Consequently, $z_1=t_\text{L}\psi_1/(t_\text{R}\psi_N)$ and $z_2=e^{i\ell \theta}$ with $\theta=2\pi/N$ ($\ell=0,\dots,N-1$), which are exactly the solutions with PBC. The eigenvalues can be obtained from the poles,  yielding $E=t_\text{R} e^{i\ell \theta}+t_\text{L} e^{-i\ell \theta}$, which is in agreement with Theorem 1(B). We obtain the coefficient of $z^m$ as $ \psi_m=\psi_N z_2^{-m}$ \cite{suppbai}. This indicates the absence of the skin effect, since $|z_2|=1$. The distribution of zeros of $P(z)$ governs the propagating factors $z_{1,2}$ and energy $E$, revealing the boundary sensitivity of non-Hermitian systems, where the eigenvalues and wavefunctions with PBC are rather different from those with OBC \cite{okuma2020topological,guo2021exact,edvardsson2022sensitivity}.

{\it (II) $\mathcal{G}(z)$ in the HN model with an impurity}:  We then discuss the formation of impurity-bound states and the associated decay rates of wavefunctions with both OBC and PBC. Based on the HN model in Eq. \ref{HNHamiltonian} of $N$ sites, we add an impurity term \cite{qi2018defect,rivero2021green,Yi2025critical,liu2019topological,bosch2019non,liu2020diagnosis,sukhachov2020non,rafi2025NHSE,het2025ocalized,moli2023anomalous,wu2023effective,guo2023accumulation,huang2025interplay,Yi2025critical}  $V a_{N_1}^\dagger a_{N_1}$.

\begin{figure}[b]
\includegraphics[width=0.49\textwidth]{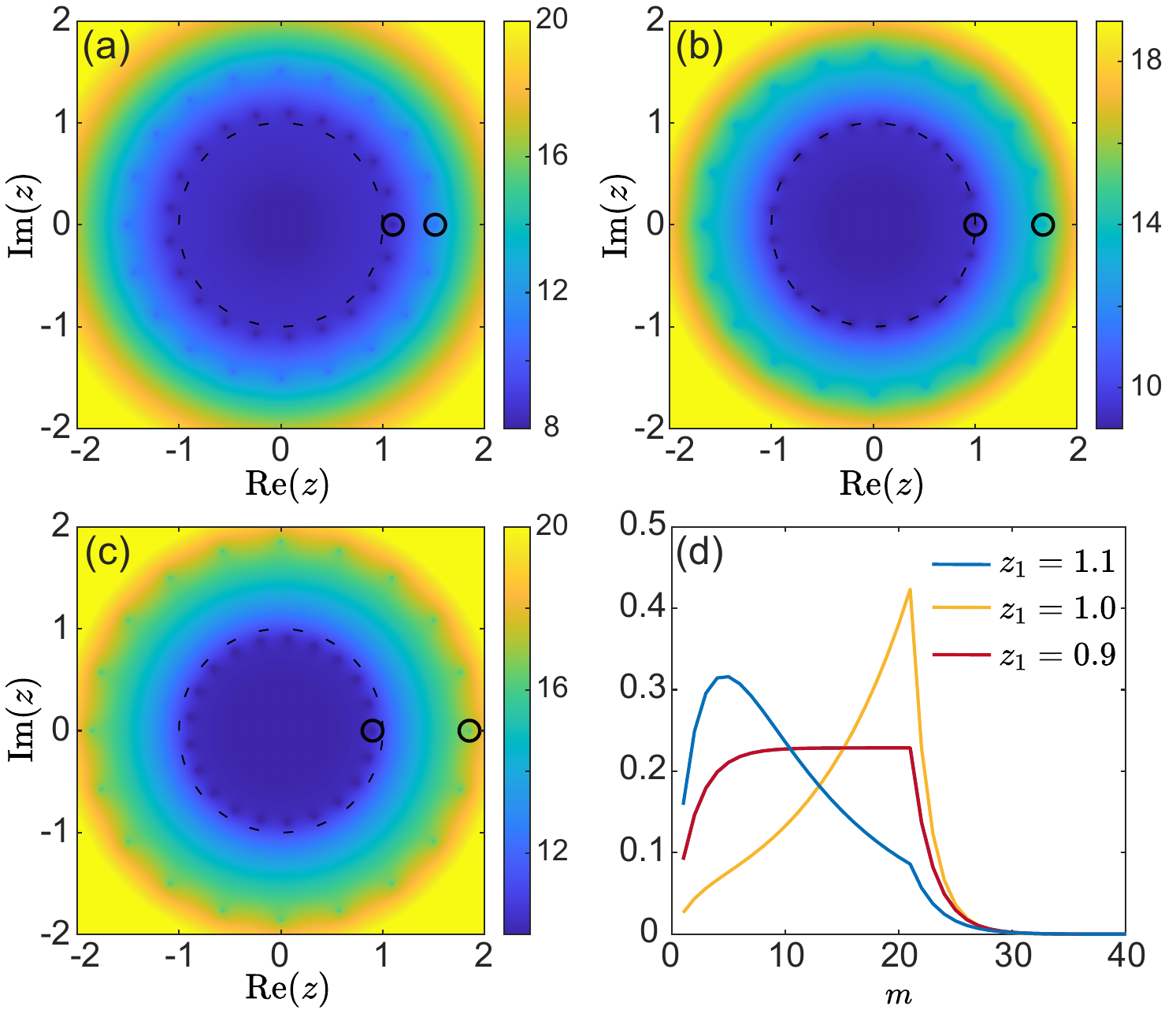}
\caption{Density plots of $\log |P(z)|$  for the impurity states of the HN model with different $V$. The black circles mark $z_1$ and $z_2$. (d) The corresponding wavefunctions localized at the left boundary, extended on the left hand side, and localized at the impurity. Parameters: (a) $V=0.2498$, (b) $V=0.4000$, and (c)  $V=0.5711$ with $N=20$, $t_\text{L}=1$, $t_\text{R}=0.6$.}
\label{fig3}
\end{figure}

(1) For OBC, with $\psi_0=0$ and $\psi_{N+1}=0$, we have
\begin{equation}
    \mathcal{G}(z)=\frac{z(t_\text{L} \psi_1+t_\text{R}\psi_N z^{N+1}-V\psi_{N_1}z^{N_1})}{t_\text{R}z^2-E z+t_\text{L}}. \label{Gimpurity}
\end{equation}
We can obtain $P(z)$ and $Q(z)$ directly from the above equation. The roots of $P(z)=0$ lie approximately on two circles with different radii [Fig. \ref{fig2}(c)]. According to Theorem 1(A),  $P(z_i)=0 $ ($i=1,2$). Following the method of derivation before, we obtain $V=V(z_1,z_2)$ \cite{suppbai}. Combined with $z_1 z_2=t_\text{L}/t_\text{R}$, it yields the condition for propagating factors with OBC in the presence of an impurity. Then the eigenvalues can be obtained by $E=t_\text{R}(z_1+z_2)$, which reproduces Theorem 1(B). We obtain the coefficients of $z^m$ as $\psi_m\propto z_1^{N_1} z_2^{N_1-m} - z_2^N z_1^{N_1-m}$ when $ m \le N_1$, and $\psi_m\propto z_1^{N+1-m} - z_2^{N+1-m}$ when $m > N_1$.

As $|V|$ increases, $|z_1|$ moves from outside to inside the unit circle while $|z_2|$ remains outside [Figs. \ref{fig3}(a)-(c)], which corresponds to the process of the wavefunction from being localized at the left boundary to being localized at the impurity [Fig. \ref{fig3}(d)]. In particular, the wavefunction is extended at the left hand side of the impurity with intermediate $V$.

(2) For PBC,  without loss of generality, we take $N_1=1$ as an example. With the condition $\psi_n=\psi_{n+N}$, we have
\begin{equation}
    \mathcal{G}(z)=\frac{z\left[(t_\text{L}\psi_1-z t_\text{R}\psi_N)(1-z^N)-V\psi_1 z\right]}{t_\text{R}z^2 -E z+t_\text{L}}. \label{GimpurityPBC}
\end{equation}
Most roots of $P(z)=0$ lie approximately on the unit circle, with the exception of two zeros off the circle [Fig. \ref{fig2}(d)].  The impurity state is obtained when $z_1$ and $z_2$ are paired with these two additional zeros \cite{suppbai}. From Theorem 1(A), $P(z_i)=0 $ ($i=1,2$). Following the method of derivation before, we obtain $V=V(z_1,z_2)$ \cite{suppbai}. Combined with $z_1 z_2=t_\text{L}/t_\text{R}$, it yields the condition for propagating factors. Then the eigenvalues can be obtained by $E=t_\text{R}(z_1+z_2)$.  We obtain the coefficients of $z^m$ as $ \psi_m = \psi_1 [z_1^{N-m+1}(z_2^N-1) - z_2^{N-m+1}(z_1^N-1)]/{(z_2^N-z_1^N)}$.

\begin{figure}[b]
\includegraphics[width=0.49\textwidth]{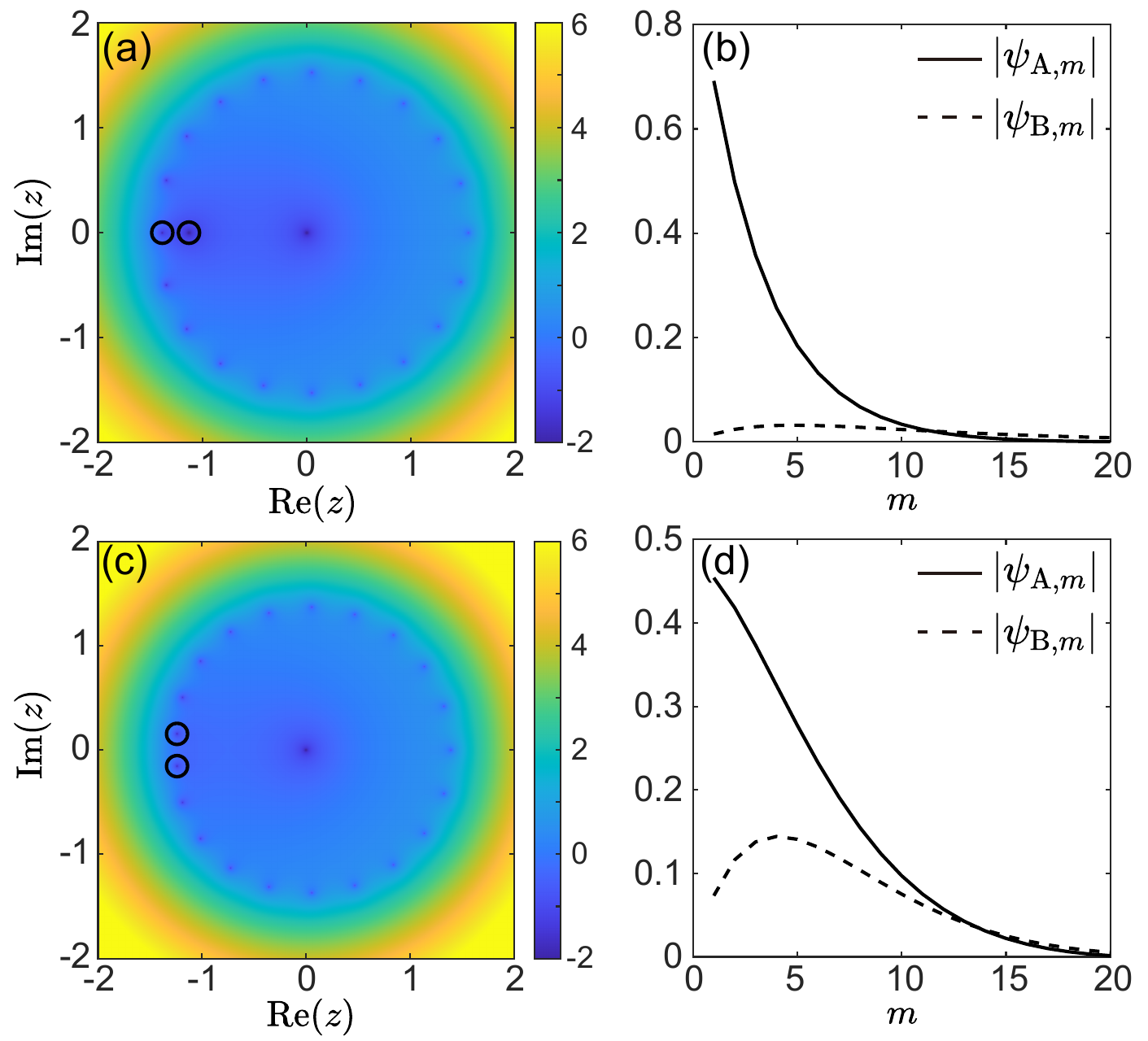}
\caption{Density plots of $\log |P_\text{A}(z)|$ for representative states of the non-Hermitian SSH model with different  $\lambda=\sqrt{t_\text{L} t_\text{R}/t'_\text{L} t'_\text{R}} $, where the black circles indicate $z_1$ and $z_2$.  The corresponding wavefunctions are shown alongside.   (a),(b) Topological edge state with $\lambda=0.9$. (c),(d) Bulk state with $\lambda=1.2$ exhibiting skin effect. Parameters are chosen as $N=20$, $t_\text{L}=1.25\lambda$, $t_\text{R}=\lambda/1.25$, $t'_\text{L}=1$, and $t'_\text{R}=1$, keeping $\sqrt{t_\text{L} t'_\text{L}/t_\text{R} t'_\text{R}}=1.25$ fixed.}
\label{fig4}
\end{figure}

{\it (III) $\mathcal{G}(z)$ in the non-Hermitian SSH model and topological phase transition}:  We now demonstrate that the above framework applies to more general physical models, particularly non-Hermitian topological models \cite{kunst2018biorthogonal,song2019non1,lee2019anatomy,hou2022deterministic}. Consider the non-Hermitian SSH chain \cite{yao2018edge,lieu2018topological,wei2025exact,lee2016anomalous,ananya2020observation,hou2022deterministic,zirnstein2021bulk} with $N$ unit cells
\begin{equation}
H=\sum_n t_\text{R} b_n^\dagger a_n+t_\text{L} a_n^\dagger b_n 
 +t'_\text{R} a_{n+1}^{\dagger} b_n+t'_\text{L} b_n^\dagger a_{n+1}.
\end{equation}
The bulk recurrence relations are given  by  $t_\text{R} \psi_{\text{A},n}+t'_\text{L} \psi_{\text{A},n+1}=E\psi_{\text{B},n}$, and 
$t'_\text{R} \psi_{\text{B},n}+t_\text{L} \psi_{\text{B},n+1}=E\psi_{\text{A},n+1}$. Define the generating functions $\mathcal{G}_{\alpha}(z)=\sum_{m=1}^{N}\psi_{\alpha,m}z^m$, where $\alpha = \text{A}, \text{B}$. Imposing OBC and using the recurrence relations, we obtain 
\begin{align}
    \mathcal{G}_\text{A}(z)&=\frac{t'_\text{L} \psi_{\text{A}, 1} z (t'_\text{R} z+t_\text{L})+t'_\text{R} E \psi_{\text{B}, N} z^{N+2}}{(t_\text{R} z+t'_\text{L})(t'_\text{R} z+t_\text{L})-E^2 z},\\
    \mathcal{G}_\text{B}(z)&=\frac{t'_\text{L} E \psi_{\text{A}, 1}z+t'_\text{R} \psi_{\text{B}, N}z^{N+1}(t_\text{R} z+t'_\text{L})}{(t_\text{R} z+t'_\text{L})(t'_\text{R} z+t_\text{L})-E^2 z}.
\end{align}
We define $\mathcal{G}_\alpha(z) = P_\alpha(z)/Q_\alpha(z)$. Remarkably,  $Q_\text{A}$ and $Q_\text{B}$ are identical. Furthermore, if we define $\boldsymbol{\mathcal{G}}(z, E)  =( \mathcal{G}_\text{A}(z, E), 
\mathcal{G}_\text{B}(z, E))^T$ and $\sigma_z=\operatorname{diag}(1,-1)$, then $\sigma_z \boldsymbol{\mathcal{G}}(z, E)  =  \boldsymbol{\mathcal{G}}(z, - E)$, which is consistent with the chiral symmetry in terms of the wavefunction in this model.
Accordingly, the propagating factors $z_{1,2}$ are identical for the $\pm E$ partners. According to Theorem 1(A),  $P_\text{A}(z_i)=0 $ ($i=1,2$), which yields
\begin{equation}
   (t'_\text{R}z_1+t_\text{L})/(t'_\text{R}z_2+t_\text{L}) =(z_1/z_2)^{N+1}. \label{x1SSH}
\end{equation}
Together with $z_1 z_2 = t_\text{L}t'_\text{L}/t_\text{R}t'_\text{R}$, this yields the propagating factors under OBC. As $N \to \infty$, the condition for the bulk states is given by $|z_1|=|z_2|= r_\text{c}$, where $r_\text{c}=\sqrt{t_\text{L}t'_\text{L}/t_\text{R}t'_\text{R}}$. We obtain the eigenvalues of this model as  $E=\pm\sqrt{t_\text{R} t'_\text{R}(z_1+z_2)+t'_\text{L} t'_\text{R}+t_\text{L}t_\text{R}}$, which reproduces Theorem 1(B). We examine how $\mathcal{G}_\alpha(z)$ governs the emergence of topological edge states from the zeros of $P(z)$ \cite{suppbai}. When $t_\text{L} t_\text{R} < t'_\text{L} t'_\text{R}$, $P(z)$ has an isolated zero of $z_1=-t_\text{L}/t'_\text{R}$, lying outside the circle of radius $(\rho/t'_\text{R})^{1/N}$. This zero corresponds to the topological edge states  \cite{yao2018edge,kunst2018biorthogonal}. Following the method of derivation before, we obtain the coefficients of $\mathcal{G}_\alpha(z)$ as $\psi_{\text{A},m}\propto (t'_\text{R} z_2+t_\text{L}) z_2^{-m}-(t'_\text{R} z_1+t_\text{L}) z_1^{-m}$ and $\psi_{\text{B},m}\propto (t_\text{R}z_1+t'_\text{L})z_1^{N-m}-(t_\text{R}z_2+t'_\text{L})z_2^{N-m}$. The variations of $z_{1,2}$ and of the wavefunctions across the topological phase transition are illustrated in Fig. \ref{fig4}.

We can characterize the topological phase transitions in the above model using the following winding number 
\begin{equation}
    \nu_\text{A}
=\frac{1}{2\pi i}\oint_{|z|=r_\text{c}} \frac{\mathrm{d}}{\mathrm{d}z}\log \mathcal{G}_\text{A}(z)|_{E=0}\,\mathrm{d}z .
\end{equation}
It characterizes whether $-t_\text{L}/t'_\text{R}$ is inside $|z| = r_\text{c}$. When $t'_\text{L}/t_\text{R}>\sqrt{t_\text{L} t'_\text{L}/t_\text{R} t'_\text{R}}$,  $\nu_\text{A}=1$;  otherwise $\nu_\text{A}=0$ \cite{suppbai}.  The same conclusion can be obtained in terms of $\mathcal{G}_\text{B}$. 

{\it (IV) Extension to two-dimensional Hermitian models}:  Finally, we show that this approach can be extended to two-dimensional (or higher-dimensional) models. Considering a one-dimensional infinite lattice with hopping $t$, the generating function satisfies $(t z+t z^{-1}-E)\mathcal{G}(z)=0$. We introduce $\mathcal{G}_\varepsilon(z)
=\sum_{m=-\infty}^{\infty}\psi_m z^m e^{-\varepsilon|m|}$ and define $\rho=e^{-\varepsilon}$. Taking a plane-wave ansatz $\psi_m=z_1^m$ with $z_1=e^{ik}$, the convergence condition of the sums over $m>0$ and $m<0$ requires $e^{-\varepsilon}<|z_1^{-1}z|<e^{\varepsilon}$. When $\rho \to 1^-$, we have $|z|\to1$, so we can set $z=e^{i\theta}$. One then obtains
\begin{equation}
      \mathcal{G}_\varepsilon(z)=\frac{1-\rho^2}{1-2\rho \cos (\theta-k)+\rho^2}.
\end{equation}
In the limit $\rho \to 1^-$,  $ \mathcal{G}_\varepsilon(z)\to 2\pi \delta(\theta-k)$, which yields $(E-2t\cos \theta)\delta(\theta-k)=0$, and hence $E=2t\cos k$.

This idea can be generalized to two dimensions. Given the equation $ E\psi_{m,n}=t(\psi_{m+1,n}+\psi_{m-1,n}+\psi_{m,n+1}+\psi_{m,n-1})$ and defining $\mathcal{G}(z_x,z_y)
= \sum_{m,n} \psi_{m,n} z_x^m z_y^n$,  we obtain $(t z_x+t z_x^{-1}+t z_y+t z_y^{-1}-E)\mathcal{G}(z_x,z_y)=0$. We introduce $ \mathcal{G}_\varepsilon(z_x,z_y)
= \sum_{m,n} \psi_{m,n} z_x^m z_y^n e^{-\varepsilon(|m|+|n|)}$
and take $\psi_{m,n}=e^{i(k_x m+k_y n)}$ with $ \rho=e^{-\varepsilon}$. When $\rho \to 1^-$, the convergence condition for the sums over positive and negative $m$, $n$ requires that $z_x=e^{i\theta_x}$, $z_y=e^{i\theta_y}$. Therefore
\begin{equation}
    \mathcal{G}_\varepsilon(z_x,z_y)  
=\prod_{\alpha=z_x,z_y}\frac{1-\rho^2}{1-2\rho\cos(\theta_\alpha-k_\alpha)+\rho^2}. 
\end{equation}
In the limit  $\rho \to 1^-$, $ \mathcal{G}_\varepsilon(z_x,z_y)\to (2\pi)^2 \delta(\theta_x-k_x)\delta(\theta_y-k_y)$,  yielding $(E-2t\cos \theta_x-2t\cos \theta_y)\delta(\theta_x-k_x)\delta(\theta_y-k_y)=0$, and hence $E=2t\cos k_x+2t\cos k_y$.  Thus, this framework yields both the spectrum and the corresponding eigenstates for Hermitian lattice models.

{\it Discussion and conclusion}: This manuscript presents a unified framework to study Hermitian and non-Hermitian models, as well as their possible  skin modes, edge states, and topological phase transitions via generating functions $\mathcal{G}(z)=P(z)/Q(z)$. We show that the zeros of $Q(z)$ should coincide with those of $P(z)$ to ensure $\mathcal{G}(z)$ 
is a finite  polynomial. The locations of these zeros determine the eigenvalues, and the coefficients of $\mathcal{G}(z)$ yield the corresponding wavefunctions. Moreover, this approach naturally leads to the GBZ condition.  We have some possible extensions of this theory to two dimensions, yielding the eigenvalues and wavefunctions simultaneously.  For these reasons, this manuscript presents a promising way to explore much intriguing physics with this new approach.

This idea suggests possible extensions to broader classes of models. Firstly, it may be extended to nonlinear non-Hermitian models.  For the Logistic map $a_{n+1} = 4 a_n(1-a_n)$ with solution $a_n = \sin^2(2^n b)$, the corresponding generating function can be written as $\mathcal{G}(z)=\frac{1}{2(1-z)}- W_{z, b}$, where $W_{z, b}$ is the Weierstrass function defined as $W_{z, b} = \sum_n z^n \cos(2^{n+1} b)/2$ \cite{Shen2018hausdorff,chaoticdynamical,peitgen2004chaos}. This map can be connected to a non-Hermitian nonlinear model with rightward hopping $r(1-a_n)$.  It is well known that the sequence $a_n$ is fully chaotic and random, highly sensitive to its initial value $b$. As a result, $\mathcal{G}(z, b)$ will become chaotic as a function of $b$, exhibiting a fractal dimension. This model provides some key insights into the possible extensions of the non-Hermitian models to nonlinear equations \cite{tobias2022fractal,mandelbrot1983fractal}. Secondly, the method can be applied to quasiperiodic models with mobility edges \cite{ganeshan2015nearest,hu2025hidden} and the associated Anderson transition. Thirdly, these physics may also be studied using the  exponential generating function \cite{chenintroduction}.  Lastly, we have checked that this idea can be generalized to time-dependent generating functions $\mathcal{G}(z,t) \equiv \sum_{m=1}^N \psi_m(t)\, z^m$ to study the dynamics of physical systems \cite{suppbai}.  

To conclude, this work  explicitly presents a direct relation between non-Hermitian models and the iterative equations in mathematics. In the future, these mathematical models could be simulated in physical experiments, given the substantial progress achieved in non-Hermitian simulations for topological phase transitions, non-Bloch bands and skin modes \cite{zou2021observation,helbig2020generalized,zhou2023observation,wang2021generating,parto2025enhanced,xiao2020non,xue2024self,zhao2025two}.





\textit{Acknowledgments}: This work is supported by the National Natural Science Foundation of China (NSFC) (Nos. T2325022, U23A2074), the Strategic Priority Research Program of the Chinese Academy of Sciences (Grant No. XDB0500000), CAS Project for Young Scientists in Basic Research (No.253 YSBR-049), Key Research and Development Program of Anhui Province (2022b1302007), Quantum Science and Technology-National Science and Technology Major Project (2021ZD0303200, 2021ZD0301200, 2021ZD0301500), and Fundamental Research Funds for the Central Universities (WK2030000107, WK2030000108). This work was partially carried out at the USTC Center for Micro and Nanoscale Research and Fabrication.

H. Y. Bai and Y. Chen contributed equally to this work.
\bibliography{ref}

\end{document}